\documentclass{article}
\usepackage[utf8]{inputenc}
\usepackage{dsfont}
\usepackage{amssymb}
\usepackage{amsmath}
\usepackage{amsthm}
\usepackage{amsfonts}
\usepackage{mathabx}
\usepackage{enumitem}
\usepackage{hyperref}
\usepackage{graphicx}
\usepackage{tabularx}
\usepackage{caption}
\usepackage{subcaption}
\usepackage[all]{xy}
\usepackage{etoolbox}

\usepackage{authblk}
\usepackage{hyperref}

\hypersetup{
    colorlinks=true,
    linkcolor=blue
    }
\begin{document}

\title{Use of Multi-CNNs for Section Analysis in Static Malware Detection}

\author{Tony Quertier\thanks{\texttt{tony.quertier@orange.com}} }
\author{Grégoire Barrué\thanks{\texttt{gregoire.barrue@orange.com}}}

\affil{Orange Innovation, Rennes, France}

\maketitle

\date{}

\begin{abstract}
Existing research on malware detection focuses almost exclusively on the detection rate. However, in some cases, it is also important to understand the results of our algorithm, or to obtain more information, such as where to investigate in the file for an analyst. In this aim, we propose a new model to analyze Portable Executable files. Our method consists in splitting the files in different sections, then transform each section into an image, in order to train convolutional neural networks to treat specifically each identified section. Then we use all these scores returned by CNNs to compute a final detection score, using models that enable us to improve our analysis of the importance of each section in the final score.
\end{abstract}

\section*{Introduction}

Static analysis is a fundamental step in malware detection, as it is the first line of defense. It provides a preliminary and quick indication of the nature of a binary file without executing it on the machine. To achieve this, there are many more or less sophisticated techniques.
Current anti-virus technologies use a signature-based approach, where a signature is a
set of rules in an attempt to identify if the binary is a malware. These rules are generally specific, and cannot usually recognize new malware so researchers have turned to artificial intelligence to improve the detection of new malware  \cite{Ucci2019, Raff2020, Gibert2020}. There are many ways of covering the subject, depending on the preprocessing chosen. For example, it is possible to learn about features extracted from binary semantic and statistical data \cite{Anderson2018}, to use language processing elements \cite{raff2018malware} or even convolutional neural networks (CNNs) \cite{marais2022ai,marais2023ameliorations}. In this article, we propose not only to improve the detection rate using multiple CNNs, but also to provide a better explainability of the results. In this aim, we use the binary-to-image Grayscale transformation proposed in \cite{Nataraj2011} and apply it to each independent section of the binary to obtain a detection score for each section. This allows us to get more information from our files, to identify the sections that appear to be the most malicious and therefore to reduce the investigation time of an analyst if necessary. Then we train a model in order to create a scoring function to predict whether the binary is malware or not using these different scores.

This article is organized as follows. In Section \ref{section1}, we give information about how we gathered our dataset composed of malware and benign binary files using Bodmas \cite{Yang} and PEMachineLearning \cite{Anderson2018} datasets, then we detail the architecture of a general PE file, and we explain our preprocessing using Grayscale method on the sections of the files. Section \ref{section2} gives the framework of our model, detailing the parameters of the algorithms that we used, while Section \ref{Results} gathers our results about the detection task and the identification of the most important features. Finally, we give a conclusion about our experiments and talk about future perspectives. 
 
\section{Dataset and preprocessing}
\label{section1}

\subsection{Dataset}

We need to prepare a reliable dataset for our experiments. In this aim we use Bodmas \cite{Yang} and PEMachineLearning \cite{Anderson2018}, and 10k recent malware from Virus Total that we only use in test. The dataset shared by Bodmas is composed of 57,293 malicious files in raw PE format, collected between August 2019 and September 2020. On the other hand, PEMachineLearning was made available by M. Lester \cite{web1}, and contains 114,737 malware among 201,549 binary files. One can find more details on these two dataset in \cite{marais2022ai}. Once all these files gathered into one dataset, we split it into three sub-datasets. The first one is used to train one CNNs per binary file's section, the second one is used to train a final scoring function, and finally the third one is kept to test the whole architecture of our algorithm. We use  80k benign 80k malware for the first sub-datasets, and the other sub-datasets are composed of 20k benign and 20k malware. For the first two sub-datasets, the train/validation ratio is $70/30$.


\subsection{PE file format}

The files that we use in our dataset are PE files (Portable Executable). It is a format used by Windows operating system that encapsulates the information necessary for the Windows OS loader to manage the wrapped executable code. The format is not architecture specific and is used to describe the structure of executable files under the Microsoft Windows family of operating systems \cite{Microsoft}. The PE format is used by the Windows system to load and execute executable files and DLLs, and is also used by many development tools, such as Microsoft Visual Studio, to create and manage executable files. The PE format is an important part of the Windows operating system and is essential for the proper functioning of many applications and system components, it is therefore a format widely used for cyberattacks, that is why we are focusing on it.

This kind of files is composed of a header and different sections. The header describes the contents of the file, such as its date of creation, its number of sections or loading information. 
Each section is described by a specific header containing its name, size and location in virtual memory. Sections generally contain the executable code (\textit{.text}) and the variables used with their default values (\textit{.data}, or in read only \textit{.rdata}). The idata section (\textit{.idata}) contains the Import Address Table (IAT), composed of libraries and their functions. The relocation section (\textit{.reloc}) contains relocation information and the resource section (\textit{.rsrc}) contains resources like icons, menus, and other elements. The resource section is a very interesting section, as it is often used by malware to evade detection. For example, scripts can be used to inject payloads directly into this section. When the binary file is executed, the embedded payload is extracted and executed. We also wanted to add the \textit{.tls} section, but we do not have enough data to exploit it, because it is not present in enough of our PE files.



\subsection{Splitting the image into sub-images}

\begin{figure}[!h]
    \centering
    \includegraphics[width=0.5\textwidth]{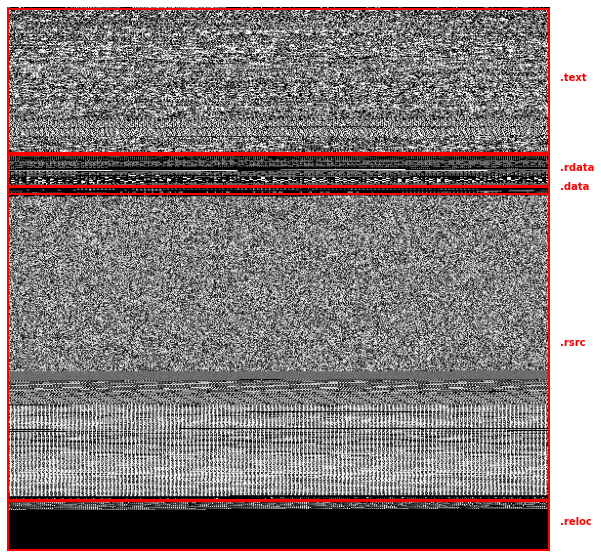}
    \caption{Sections in a malware}
    \label{fig:sections}
\end{figure}

\begin{figure}[!h]
    \centering
    \begin{subfigure}[h]{0.2\textwidth}
        \includegraphics[width=\textwidth]{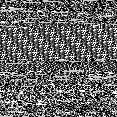}
        \caption{example of a section \textit{.text}}
    \end{subfigure}
    \hspace{1cm}
    \begin{subfigure}[h]{0.2\textwidth}
        \includegraphics[width=\textwidth]{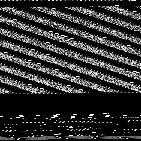}
        \caption{example of a section \textit{.data}}
    \end{subfigure}
    \hspace{1cm}
        \begin{subfigure}[h]{0.2\textwidth}
        \includegraphics[width=\textwidth]{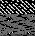}
        \caption{example of a section \textit{.idata}}
    \end{subfigure}

        \begin{subfigure}[h]{0.2\textwidth}
        \includegraphics[width=\textwidth]{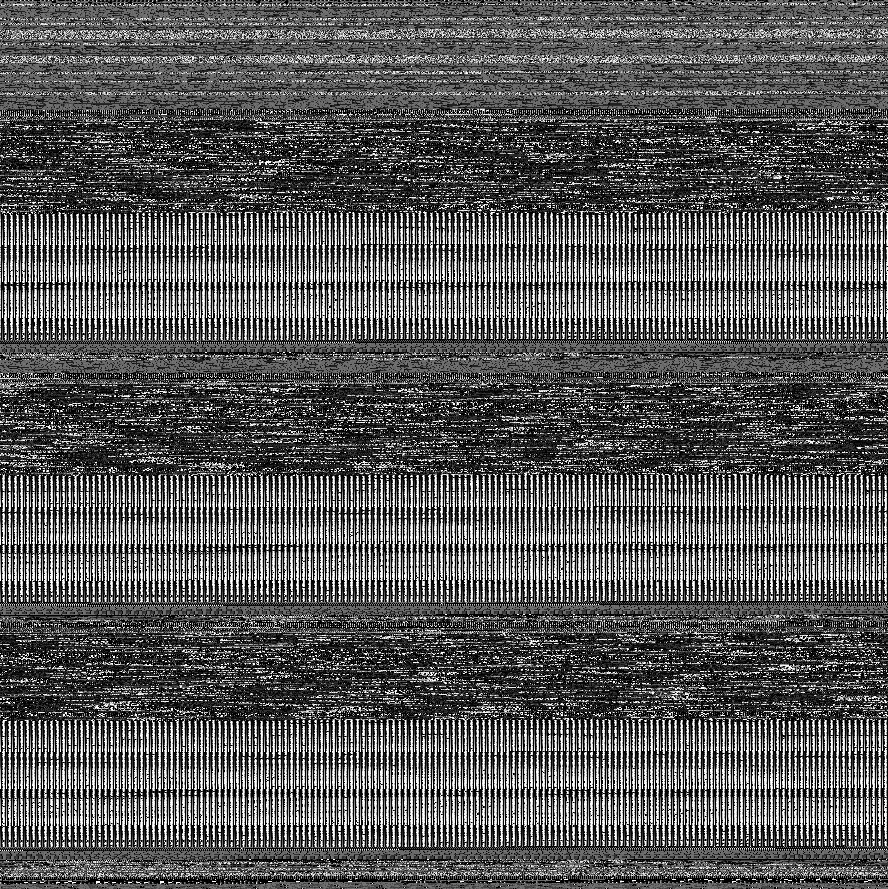}
        \caption{example of a section \textit{.rdata}}
    \end{subfigure}
    \hspace{1cm}
    \begin{subfigure}[h]{0.2\textwidth}
        \includegraphics[width=\textwidth]{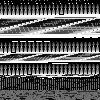}
        \caption{example of a section \textit{.rsrc}}
    \end{subfigure}
    \hspace{1cm}
    \begin{subfigure}[h]{0.2\textwidth}
        \includegraphics[width=\textwidth]{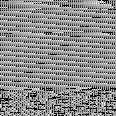}
        \caption{example of a section \textit{.reloc}}
    \end{subfigure}
    \caption{Images of different binary's sections }
    \label{fig:sections_découpées}
\end{figure}

The preprocessing of our data consists in applying the Grayscale method \cite{Nataraj2011}, which transforms a file into an image, on each different section of our files. These sections are identified in the files using the LIEF library \cite{LIEF}, and then are transformed into $64\times64$ images. Here we have considered 7 sections as we think they are relevant, and focus on these when they are present. In Figure \ref{fig:sections} one can see the result of Grayscale method, resulting in an image corresponding to a PE file. On this image we separate with red lights the different sections that compose the file, whereas in Figure \ref{fig:sections_découpées} we present the Grayscale method used on several sections of the file, giving different images.


\section{Framework}
\label{section2}

Once our preprocessing done, we get sub-datasets for each section that we considered. For each section we separately train a CNN. Once the parameters of each CNN have been set, we use the second dataset, and assign to each file a vector containing the scores (corresponding to the probability to be classified as a malware) of each section. If some sections are not present in the file, we give these sections a score of $-1$. Initially we had chosen a neutral score of $0.5$, but this introduced a bias because this is equivalent to considering that the absence of a section does not matter. However, the presence or absence of certain sections can provide additional information about a binary file. Thus we train 7 CNNs, corresponding to the 7 sections that we identified  for our experiments. As the \textit{.tls} section was underrepresented, we chose not to consider it to keep working with comparable sizes of training. Each CNN is built with three layers, a dropout rate of 0.2, and for the training phase we use batch sizes of 64 and  a maximum of 100 epochs using early stopping.

\begin{figure}[!h]
    \centering
    \includegraphics[width=1\textwidth]{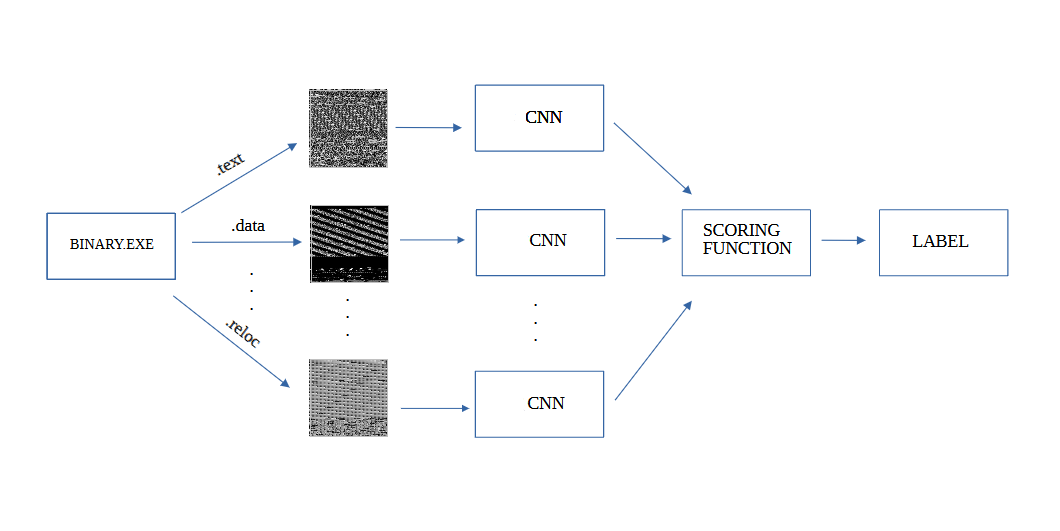}
    \caption{Architecture of our algorithm. We start from a binary file, we decompose it into images of its sections. Each section is used to train a specific CNN, then we gather the scores of the CNN in a vector used as input to train a scoring function. Once everything is trained, we get a model which take the binary as input and give its predicted label as output. }
    \label{fig:archi}
\end{figure}

As an output of this step, we therefore have a vector composed of the six scores, and we want a classifying decision as a final output.  Hence, we need to find a customized scoring function to give a final answer to this classifying task given all the scores coming from the CNNs. One could think about using a majority vote, but this would include a bias that suggests that every section is equally important. We test various functions and train some algorithms to try and find the most optimal function for calculating this score. For the final version of the algorithm, we keep a XGBoost  and a RF model, which gave us the best results. Figure \ref{fig:archi} gives a good overview of our entire model, while the various experiments and results can be found in section \ref{Results}.

\section{Experiments and results}
\label{Results}

In this section we present the results of our algorithm, which contains three steps. First, we train a CNN model for each section of the files. When a section is not present in the file, there is no image corresponding to this section, so the different CNNs are not trained on exactly the same sizes of dataset, but this guarantees the unbiased property of the training. For instance, on the \textit{.idata} section, the training and the test are made on less data, because this section is present in only 11,000 benign files and 6,000 malware files. It is even worse for the \textit{.tls} section, present in only 3,400 benign files and 1,000 malware files.  

Table \ref{tab:section} gathers the training and test accuracy for each section. Since the rates of false-positive and false-negative are very important for our use-case (where sometimes data could be unbalanced) we also compute the F1-score, which gives an additional proof of the performances of our results. The results for the different sections are quite similar, the \textit{.rsrc} or \textit{.idata} sections getting slightly better accuracy and F1-score. This could highlight the importance of taking these sections into account for the detection task. It is not very surprising that these sections are determining because they contain lots of information about the files. For example the \textit{.idata} gathers the imports of the file, and we know that these features are very impacting in this use-case \cite{marais2023ameliorations}. 



\begin{table}[!ht]
        \caption{CNN training results for each section }
        \label{tab:section}
   \centering
    \begin{tabular}{|c|c|c|c|c|c|}
    \hline
      & Train & Valid & \multicolumn{2}{c|}{Test} \\
    \hline
     Section & Accuracy & Accuracy & Accuracy & F1-score \\
    \hline
    text & 0.98 & 0.92 & 0.92 &  0.84 \\
    \hline
    data & 0.98 & 0.92 & 0.93 & 0.87 \\
    \hline
    rdata & 0.98 & 0.93 & 0.92 & 0.89 \\
    \hline
    rsrc & 0.98 &0.94 & 0.94 & 0.90 \\
    \hline
    reloc & 0.98 & 0.93 & 0.94 & 0.80 \\
    \hline
    idata & 0.99 & 0.97 & 0.97 & 0.94 \\
    \hline
    tls & 0.85 & 0.84 & 0.84 & 0.5 \\
    \hline
    \end{tabular}
\end{table}

Once the CNNs are trained, we use them with the second part of the dataset in order to assign to each PE file a vector containing the scores associated to each section. Then we have to train the scoring function on this dataset. In order to identify the optimal scoring functions for our problem, we try several models, which are presented in Table \ref{tab:scoring_function}. We can see that the XGBoost and the Random Forests (RF) models have the best performances, the RF model being better on the train but comparable on the test. The LightGBM model performs slightly worse than XGBoost on the train, so we do not keep it for the final step. Note that we also try a majority vote, which as expected is not relevant in our use-case.

\begin{table}[!ht]
        \caption{Scoring function training results}
        \label{tab:scoring_function}
   \centering
    \begin{tabular}{|c|c|c|c|c|c|}
    \hline
      & \multicolumn{2}{c|}{Train} & \multicolumn{2}{c|}{Test} \\
    \hline
     Section & Accuracy & F1-score & Accuracy & F1-score \\
    \hline
    XGBoost & 0.99 & 0.99 &  0.98 & 0.98 \\
    \hline
    LGBM & 0.987 & 0.987 & 0.98 & 0.98 \\
    \hline
    RF & 0.99 & 0.99 & 0.98 & 0.98 \\
    \hline
     Maj. Vote $\geq 3$ &  &  & 0.86 & 0.85 \\
    \hline
    Maj. Vote $> 3$ &  &  & 0.76 & 0.72 \\
    \hline
    \end{tabular}
\end{table}

\begin{table}[!ht]
    \begin{center}
        \caption{Testing multi-CNN with two different scoring function on third dataset}
        \label{tab:test}
        \begin{tabular}{|c|c|c|}
            \hline
             &  Accuracy & F1-score \\ 
             \hline
             RF  &  0.96 & 0.96 \\ 
             \hline
             XGBoost & 0.96 & 0.96 \\
             \hline
        \end{tabular}
    \end{center}
\end{table}
Finally, once both the CNNs and the scoring function are trained, we use the last part of the dataset as a test sample, to evaluate the performances of the whole model. We compare two models in Table \ref{tab:test}, which give the same results. For comparison, in paper \cite{marais2022ai} on a similar dataset, we have 0.945 in accuracy and 0.95 in F1-score. For accuracy, we gain 1.5 $\%$, which may not seem like a huge gain, but in cybersecurity it's not insignificant especially as we had to split our dataset to train the different bricks of our algorithm, thus the models are trained on less data. In addition, this improvement also comes with better interpretability of the results and potential enhancement by adding sections or customizing the various CNNs.
Note that Random Forests allow more explicability, as we can identify the decisions made by the algorithms and understand the importance of some particular features.

\begin{figure}[!h]
    \centering
    \begin{subfigure}[h]{0.47\textwidth}
        \includegraphics[width=\textwidth]{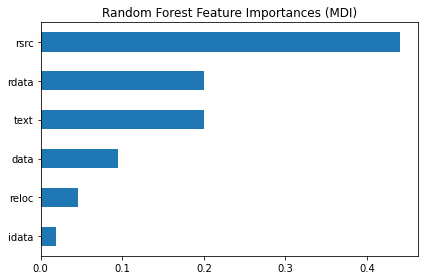}
        \caption{MDI for Random Forest}
    \end{subfigure}
    \hspace{0.3cm}
    \begin{subfigure}[h]{0.47\textwidth}
        \includegraphics[width=\textwidth]{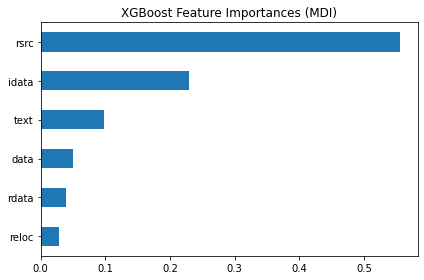}
        \caption{MDI for XGBoost}
    \end{subfigure}
    \caption{Mean Decrease Impurity  for RF and XGBoost models}
    \label{fig:MDI}
\end{figure}

\begin{figure}[!h]
    \centering
    \begin{subfigure}[h]{0.47 \textwidth}
        \includegraphics[width=\textwidth]{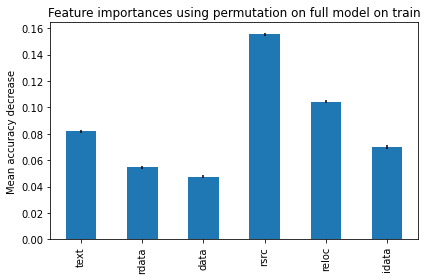}
        \caption{permutation feature on training model of RF}
    \end{subfigure}
    \hspace{0.5cm}
    \begin{subfigure}[h]{0.47 \textwidth}
        \includegraphics[width=\textwidth]{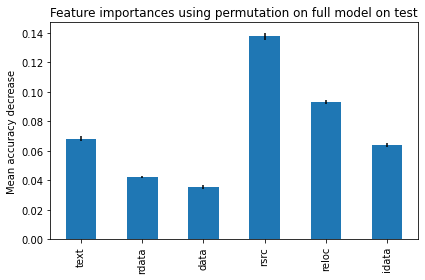}
        \caption{permutation feature on test model of RF}
    \end{subfigure}

    \begin{subfigure}[h]{0.47 \textwidth}
        \includegraphics[width=\textwidth]{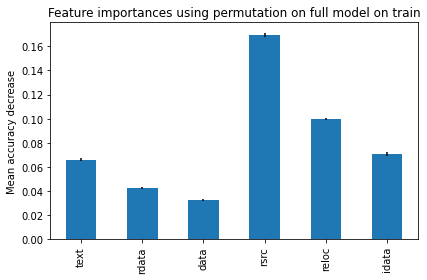}
        \caption{permutation feature on training model of XGBoost}
    \end{subfigure}
    \hspace{0.5cm}
    \begin{subfigure}[h]{0.47 \textwidth}
        \includegraphics[width=\textwidth]{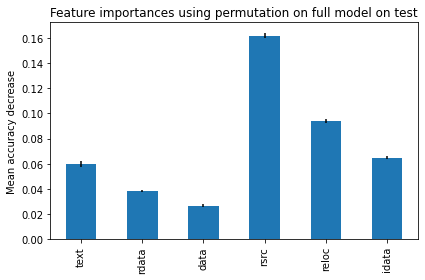}
        \caption{permutation feature on test model of XGBoost}
    \end{subfigure}
    \caption{Permutation feature importance for XGBoost and RF models}
    \label{fig:MDA}
\end{figure}

In order to  evaluate the impact of the different sections in the scoring of our binary files, we compute Mean Decrease in Impurity (MDI), presented on Figure \ref{fig:MDI} and Feature permutation, presented on Figure \ref{fig:MDA}. Mean Decrease in Impurity calculates each feature importance as the sum over the number of splits (across all trees) that include the feature, proportionally to the number of samples it splits.

As said in Scikit-learn documentation, the impurity-based feature importance of Random Forests suffers from being computed on statistics derived from the training dataset and therefore do not reflect the ability of feature to be useful to  predictions that generalize to the test set (when the model has enough capacity). Besides the impurity-based importance is biased towards high cardinality.



Permutation-based feature importance do not exhibit such a bias. Additionally, the permutation feature importance may be computed with any performance metric on the model predictions and can be used to analyze any model class. Random Forests and other tree-based algorithm use the OOB (out-of-bag) samples to construct a different feature-importance measure for the prediction strength of each feature. When the tree is grown, the OOB samples are passed down the tree, and the prediction accuracy is recorded. Then the values for the focused feature are randomly permuted in the OOB samples, and the accuracy is again computed. The decrease in accuracy as a result of this permuting is averaged over all trees, and is used as a measure of the importance of the focused feature in the Random Forest \cite{hastie2009elements}.

A number of things can be deduced from these results. Firstly, it ensures that there was no overfitting of our model during the training phase, since in Figure \ref{fig:MDA} we do not observe any change in the permutation feature importance between the training and the test phases. By looking only at the scores of Table \ref{tab:test}, we do not have any element to know what model is the best. Thus, we need to identify what features are the more relevant for the classification task, and then make sure that our model take into account the importance of these features. The feature permutation phase allows to highlight the importance of the \textit{.idata} and \textit{.rsrc} sections. When looking at the MDI in Figure \ref{fig:MDI}, we can see that the RF model does not gives a lot of importance to the \textit{.idata} section, whereas it is considered as the second most important section in the training set for the XGBoost model. This indicates that the XGBoost model is more suitable for our use-case. This approach could be useful to add more sections in the classification task: even if they are underrepresented, the permutation feature importance will tell us about the impact of these new features, while the MDI will ensure whether or not our model take them into account.  



\section{Conclusion and future work}

The power of our proposed algorithm is that we can explore many different directions to enhance its performances. First we can increase the number of sections taken into account (\textit{.edata}, \textit{.bss}, ...), in order to understand their impact on the final score. This could be done very efficiently as we do not have to re-train the CNNs on the already studied sections, but just train some new CNNs on the additional sections and take them into account in the input of the scoring function. We can also investigate what scoring function would be the most relevant for our problem. The use of Random Forests or XGBoost allows to get information about the sections, and thus to identify which sections are the most likely to help for the classification task. We can see, for example, that although the \textit{.idata} section is rarely present, when it is, it provides a great source of information. So we would like to highlight that our approach gives us both a detection score and the sections that hold the most information for an analyst. As we study the impact of this section and its limited presence, we are encouraged to add more sections, even if they are not as regularly present as the standard ones, because we can quantify their impact using permutation feature importance, and make sure that our model learns correctly from them with the MDI.
Besides, once this information gathered, we could implement a weighted average scoring function to give a percentage of maliciousness of the files, hence giving a more nuanced response to the problem. 

In conclusion, we proposed in this work a distributed CNN model in order to classify malware and benign files. We identified the different sections of these files, and transformed each section into an image thanks to Grayscale method. Then, we selected six specific sections, and trained a CNN on each section. Once the training done, we gathered the six CNNs outputs into vectors, that we studied using a scoring function in order to identify the type of each file. We also performed tests to identify which sections are the most impacting for our use-case and to verify that our model take them into account. Our results for different scoring functions show the efficiency of this model compared to a CNN classifying images of entire files. Besides, the structure of this model makes it easy to modify, adapt, and allows to get more information about the data, increasing our understanding about malware PE files.

\bibliographystyle{unsrt}
\bibliography{ref}

\end{document}